\documentclass[twocolumn,showpacs,showkeys,prd,superscriptaddress]{revtex4-2}

\usepackage{amsmath,amssymb,amsfonts,dsfont,mathrsfs,amsthm}
\usepackage{graphicx}
\usepackage{centernot}
\usepackage{hyperref}
\usepackage{xcolor}
\usepackage{comment}
\hypersetup{linktocpage,colorlinks=true,urlcolor=blue!80!red,linkcolor=blue,citecolor=red}
\usepackage{feynmf}
\usepackage{siunitx}
\usepackage{array}
\usepackage{ulem}
\usepackage{tikz}
\usepackage{braket}
\usepackage{float}

\begin{document}

\title{Predictive power of the Berezinskii-Kosterlitz-Thouless theory based on Renormalization Group throughout the BCS-BEC crossover in 2D superconductors}

\author{Giovanni \surname{Midei}}
\affiliation{School of Science and Technology, Physics Division, University of Camerino,
Via Madonna delle Carceri, 9B, 62032 - Camerino (MC), Italy}
\affiliation{Istituto Nazionale di Fisica Nucleare, Sezione di Perugia, Via A. Pascoli, 23c, 06123 Perugia, Italy}
\author{Koichiro \surname{Furutani}}
\affiliation{Dipartimento di Fisica e Astronomia ‘Galileo Galilei’ and QTech Center, Università di Padova, Via Marzolo 8, 35131 Padova, Italy}
\affiliation{Istituto Nazionale di Fisica Nucleare, Sezione di Padova, Via Marzolo 8, 35131 Padova, Italy}
\author{ Luca \surname{Salasnich}}
\affiliation{Dipartimento di Fisica e Astronomia ‘Galileo Galilei’ and QTech Center, Università di Padova, Via Marzolo 8, 35131 Padova, Italy}
\affiliation{Istituto Nazionale di Fisica Nucleare, Sezione di Padova, Via Marzolo 8, 35131 Padova, Italy}
\affiliation{Istituto Nazionale di Ottica, Consiglio Nazionale delle Ricerche, Via Carrara 2, 50019 Sesto Fiorentino, Italy}
\author{Andrea \surname{Perali}}
\affiliation{School of Pharmacy, Physics Unit, University of Camerino, Via Madonna delle Carceri, 9B, 62032 - Camerino (MC), Italy}

\begin{abstract}
Recent experiments on 2D superconductors allow the characterization of the critical temperature and of the phase diagram across the BCS-BEC crossover as a function of density. We obtain from these experiments the microscopic parameters of the superconducting state at low temperatures by the BCS mean-field approach. For Li$_x$ZrNCl, the extracted parameters are used to evaluate the superconducting phase stiffness and the Berezinskii-Kosterlitz-Thouless (BKT) critical temperature throughout the BCS-BEC crossover, by implementing the corresponding Renormalization Group (RG) approach. In this way, we make a quantitative test of the predictive power of the BKT theory for evaluating the critical temperature. The RG flow equations turn out to give a sizable renormalization of the phase stiffness and of the critical temperature, which is crucial to obtain a satisfactory agreement between the BKT theory and the experiments, in particular in the BCS-BEC crossover regime. We predict the temperature range where phase stiffness renormalization can be measured in Li$_x$ZrNCl across the BCS-BEC crossover. Contrary to other microscopic theories of superconductivity, we find that the BKT theory can be exploited to evaluate quantitatively the critical temperature of 2D superconductors in different pairing regimes.  
\end{abstract}

\maketitle

\textit{Introduction.}---Two-dimensional superconductivity in ultra-thin materials has been recently achieved in several systems, with precise control of superconducting properties by tuning the carrier density or the geometry of the samples \cite{Zhang,Farrar,Schneider,Nakagawa, Randeria}. At low temperature the BCS-BEC crossover has been investigated \cite{Uemura,Shibauchi}, while at finite temperature the BKT topological transition has been characterized experimentally by detection of the jump of the phase stiffness at the BKT critical temperature $T_{BKT}$ and from I-V characteristics \cite{Sharma, Venditti}. In the case of 2D ultracold atomic gases, comparisons between theory and experiments for $T_{BKT}$ throughout the BCS-BEC crossover have been reported \cite{Yao, Sobirey, Bighin, Tempere, Hadzibabic, Dogra, Kruger, Bloch, Botelho, Lin}. On the other hand, a quantitative comparison between the RG-based BKT theory and experimental data for $T_{BKT}$ across the BCS-BEC crossover is lacking for solid state systems, which is the aim of this work. Recent experimental characterization of $T_{BKT}$ across the BCS-BEC crossover in  Li$_x$ZrNCl, upon tuning Lithium concentration by gating, has been reported in Ref. \cite{Nakagawa}. The data for the superconducting state quantities as a function of doping allow us to extract the microscopic parameters characterizing the pairing interaction and the chemical potential of the system. Specifically, by inserting as input the low temperature superconducting gap and the carrier density, we obtain the pairing interaction and the chemical potential by solving the BCS mean-field equations. 
The superconducting gap close to $T=0$ is the most suitable quantity of the
superconducting state to extract the pairing parameters by the BCS theory 
because of the negligible effects of fluctuations away from the critical temperature. With the obtained microscopic parameters, cross checked with the available data for the phase and Pippard coherence lengths, we study the BKT transition without free parameters and evaluate $T_{BKT}$ as a function of doping, with and without the solution of the RG flow equations. We find that the BKT theory has a remarkable predictive power against experimental doping 
dependence of the critical temperature and that the RG corrections are necessary for the best comparison
with the experimental data, including the peculiar inflection point not well reproduced in other theoretical approaches \cite{Shi}. 

Remarkably, very recently a striking proof of a BKT superconducting transition has been reported in Ref. \cite{Weitzel} in strongly disordered ultra-thin NbN films. Measurements show that the temperature dependence of the phase stiffness presents a sharp jump at $T_{BKT}$ with a quantitative overlap between theory and experiment. This jump occurs in a window of temperature that is even sharper than the one observed in ultracold gases \cite{Noh}. Wietzel at al. \cite{Weitzel} show that at temperatures much smaller than $T_{BKT}$, the phase stiffness follows the behavior predicted by the BCS theory, a well known feature both for ultracold gases \cite{Bighin} and solid state superconductors \cite{lara}. However, as $T_{BKT}$ is approached, in a range of $50$ mK the BCS phase stiffness deviates from the renormalized stiffness of the RG theory, in perfect agreement with experiment. Even in a weak-coupling superconductor such as NbN the deviation from the BCS behavior due to the RG effects, that are expected to be small, can be observed. This provides additional motivations to explore the predictive power of the RG for $T_{BKT}$ in the entire crossover. We predict that the deviations of the phase stiffness from the BCS behavior becomes larger in the BCS-BEC crossover regime and, hence, the RG becomes crucial for correctly reproducing the temperature dependence of the phase stiffness and thus the quantitative value of $T_{BKT}$ observed in the experiments \cite{Nakagawa}. These deviations can be observed in Li$_x$ZrNCl where the BCS-BEC crossover can be explored by tuning the Lithium concentration, in contrast with NbN which is a weakly-coupled BCS superconductor.\\

\textit{Model and methods.}---We consider two degenerate parabolic bands with dispersion $\varepsilon(\mathbf{k})=\mathbf{k}^2/2m^*$ to mimic the single band of Li$_x$ZrNCl having two valleys and $m^*=0.9m_e$. The interactions are equal in both bands and the two bands are independent. We set $k_B, \hbar=1$ throughout the paper. In Ref.\cite{Nakagawa} the Cooper pairing has been attributed to an electron-phonon mechanism. Thus, we model the two-particle interaction with a separable potential $V(\mathbf{k}, \mathbf{k'})=
V(\mathbf{k}, \mathbf{k'})= -V_{0} w(\mathbf{k})w(\mathbf{k'})$, where $\omega_0$ is the energy cutoff, $V_0>0$ is the strength of the pairing potential and
\begin{equation}
w(\mathbf{k})=\frac{1}{2}\Bigl[\tanh{\Bigl(\frac{\xi(\mathbf{k})+\omega_0}{2t_{eff}}\Bigr)}-\tanh{\Bigl(\frac{\xi(\mathbf{k})-\omega_0}{2t_{eff}}\Bigr)}\Bigr]
\label{eqn:1111}
\end{equation}
where $\xi(\mathbf{k})=\varepsilon(\mathbf{k})-\mu$ is the band dispersion measured from the chemical potential $\mu$. The function $w(\mathbf{k})$ is a smooth approximation of the Heaviside function in the BCS theory and its width is determined by the effective energy scale $t_{eff}$ in order to obtain finite derivatives of the potential, essential to get the coherence lengths. We fix a small $t_{eff}=0.3$ meV to keep the step-like feature of the BCS-like pairing, while different values of $\omega_0$ are considered.
The dimensionless coupling parameter is $\lambda=N V_0$, where $N=m^*/ \pi$ is the 2D density of states per band.
We study the superconducting state of the system using the BCS  equation 
\begin{equation}
\Delta(\mathbf{k})=-\frac{1}{2\Omega} \sum_{k'}V(\mathbf{k}, \mathbf{k'})\frac{\Delta(\mathbf{k'})}{E(\mathbf{k'})} \tanh{\frac{E(\mathbf{k'})}{2 T}}
\label{eqn:14}
\end{equation}
where $\Delta(\mathbf{k})$ is the superconducting gap and $E(\mathbf{k})=\sqrt{{\xi(\mathbf{k})}^2+{\Delta(\mathbf{k})}^2}$ is the dispersion of single-particle excitations. Eq. (\ref{eqn:14}) has to be coupled with the equation for the total density,
\begin{equation}
n=\frac{4}{\Omega}  \sum_{k} \Bigl[{v(\mathbf{k})}^2 f\big(-E(\mathbf{k})\big)+{u(\mathbf{k})}^2 f\big(E(\mathbf{k})\big)\Bigr]
\label{eqn:17}
\end{equation}
where $f(E)$ is the Fermi-Dirac distribution, $u(\mathbf{k})$ and $v(\mathbf{k})$ are the BCS coherence weights. The factor $4$ takes into account the spin degeneracy and the two valleys in the conduction band of Li$_x$ZrNCl. We extrapolate the value of $\Delta$ and of the Fermi energy $E_F$ from the data in Ref. \cite{Nakagawa} and calculate $n$ corresponding to a given Li concentration $x$. Then, we solve the zero temperature limit of Eqs.(\ref{eqn:14}) and (\ref{eqn:17}) to evaluate $\lambda$ and $\mu$. Since the system is in a 2D regime for all the considered values of $x$, the superconducting transition is described by the BKT physics. Once we have calculated the microscopic parameter of the superconducting ground state, we calculate the superconducting phase stiffness $J$ and $T_{BKT}$.
To evaluate $J$, we follow the approach used in Appendix A of Ref. \cite{Metzner}, considering the response function $\chi_{\alpha \alpha'}(\mathbf{q}, \omega)$ of the system to an external electromagnetic field. The index $\alpha$ refers to the Cartesian axis. $J$ is related to the static and long-wave-length limit of the response function $\chi_{ \alpha \alpha'}=\lim_{\mathbf{q}\to 0} \chi_{ \alpha \alpha'}(\mathbf{q},0)$.
$\chi_{ \alpha \alpha'}$ is given by the sum of a paramagnetic and a diamagnetic contribution, that is, $\chi_{ \alpha \alpha'}=\chi_{ \alpha \alpha'}^{para}+\chi_{ \alpha \alpha'}^{dia}$, where
\begin{equation}
 \chi_{ \alpha \alpha'}^{para}= \frac{2e^2}{V} \sum_{\mathbf{k}} f'\big(E(\mathbf{k})\big) \varepsilon^{\alpha}(\mathbf{k}) \varepsilon^{ \alpha'}(\mathbf{k})
\label{eqn:23}
\end{equation}
\begin{equation}
 \chi_{ \alpha \alpha'}^{dia}= \frac{e^2}{V} \sum_{\mathbf{k}} \Biggl[ 1-\frac{\xi(\mathbf{k})}{E(\mathbf{k})}+\frac{2\xi(\mathbf{k})}{E(\mathbf{k})}f\big(E(\mathbf{k})\big) \Biggr] \varepsilon^{\alpha \alpha'}(\mathbf{k})
\label{eqn:24}
\end{equation}
with $\varepsilon^{ \alpha}(\mathbf{k}) = \partial{\varepsilon (\mathbf{k}})/\partial{k_{\alpha}}$, and $\varepsilon^{\alpha \alpha'}(\mathbf{k}) = \partial^2{\varepsilon (\mathbf{k})}/ \bigl(\partial{k_{\alpha}}\partial{k_{\alpha'}}\bigr)$
The off-diagonal elements ($\alpha \neq \alpha'$) of the response function vanish since the dispersion is symmetric in $k_x$ and $k_y$. Furthermore, the response function does not depends on the direction ($\chi_{ \alpha}= \chi$) and $J$ is given by 
\begin{equation}
J=\chi/(2e)^2.
\label{eqn:1000000}
\end{equation}
As a first approximation, $T_{BKT}$ is given by the Nelson-Kosterlitz (NK) criterion
\begin{equation}
T_{BKT}= \frac{\pi}{2} J (T_{BKT}).
\label{eqn:100}
\end{equation}
In order to take into account the band structure of Li$_x$ZrNCl, we consider a contribution $j$ to the stiffness for each valley, so that the stiffness of the system is given by $J=2j$.
The NK criterion in Eq.(\ref{eqn:100}) provides the temperature at which the presence of a single vortex in the superconducting phase becomes energetically favorable. However, in 2D real systems the superconducting phase is characterized by the presence of a ‘gas’ of vortex-antivortex pairs and $T_{BKT}$ is the temperature at which the vortex-antivortex pairs unbind and the quasi-ordered phase is lost. The presence of multiple topological defects leads to a suppression of $T_{BKT}$ obtained using the approximation of Eq.(\ref{eqn:100}). The presence of vortex-antivortex pairs in the low temperature phase can be taken into account through an RG approach.
As shown by Kosterlitz and Thouless \cite{Kosterlitz1977}, the RG flow equations are
\begin{subequations}
\begin{equation}
 \frac{dK^{-1}_l(T)}{dl}=4 \pi^3 y_l^2(T)
\label{eqn:3000}
\end{equation}
\begin{equation}
\frac{dy_l(T)}{dl}=(2-\pi K_l(T))y_l(T)
\label{eqn1}
\end{equation}
\end{subequations}
where $l$ is the scaling parameter, which goes from $l = 0$ (bare results) to $l = \infty$ (fully renormalized
results), $K_l(T)= \beta J_l(T)$, with $\beta=1/ T$, $y_l(T)= \exp{(-\beta \mu_{vor,l})}$ is the vortex fugacity, and $\mu_{vor,l}$ is the vortex core energy. According to our notation $J_{l=0}(T)$ corresponds to the bare phase stiffness given by Eq.(\ref{eqn:1000000}).
The system of differential equations (Eqs.(\ref{eqn:3000}) and (\ref{eqn1})) can be solved by separation of variables \cite{Stoof} and it yields
\begin{equation}
    y_l(T)^2-\frac{1}{2 \pi^3}\Bigl[\frac{2}{K_l(T)}+ \pi \ln{K_l(T)}\Bigr]=C
\label{1c}
\end{equation}
where $C$ is an integration constant determined by the initial conditions. On the critical trajectory, where $T = T_{BKT}$, one has $y_{\infty}(T_{BKT})=0$ and $K_{\infty}(T_{BKT}) = 2/\pi,$ the renormalized critical temperature $T_{BKT}$ is then implicitly defined by using the NK criterion in Eq.(\ref{eqn:100}), replacing the bare phase stiffness with the renormalized one
\begin{equation}
    T_{BKT}=\frac{\pi}{2}J_{\infty} (T_{BKT}).
\label{1t}
\end{equation}

Note that Eq.(\ref{1t}) relates $T_{BKT}$ to the fully renormalized phase stiffiness $J_{\infty} (T_{BKT})$ calculated at $T_{BKT}$. However, it is possible to derive an expression for $T_{BKT}$ that depends only on the bare phase stiffness $J_0$.
In fact, by taking into account the definitions of $y_0(T)$ and $K_0(T)$, we have for the vortex fugacity
\begin{equation}
 y_0(T)=  e^{-\frac{\pi^2}{4} K_0(T)}.
\label{ss}
\end{equation}
In order to obtain Eq.(\ref{ss}), we have chosen as initial conditions for the vortex core energy $\mu_{vor,0}(T)=\frac{\pi ^2}{4}J_0(T)$ that is currently the best choice for superconductors and superfluids \cite{Minnhagen, Khawaja, Zhang2008, Bighin}. Thus from Eqs.(\ref{1c}) and (\ref{ss}) evaluated at $T_{BKT}$ we have
\begin{equation}
  \begin{split}
   e^{-\frac{\pi^2}{2} K_0(T_{BKT})}-\frac{1}{2 \pi^3}\Bigl[\frac{2}{K_0(T_{BKT})}\\
   + \pi \ln{K_0(T_{BKT})}\Bigr]=-0.02778
\label{pp}
  \end{split}
\end{equation}
using the value $C=-0.02778$ for the integration constant that identifies the critical trajectory \cite{Stoof}. The solution of Eq.(\ref{pp}) is $K_0(T_{BKT})=  1.055$. Since $K_0(T_{BKT})= J_0(T_{BKT})/T_{BKT}$, we obtain
\begin{equation}
 T_{BKT}=0.948 J_0(T_{BKT}).
\label{bb}
\end{equation}
This equation can be seen as a renormalized NK criterion, in which the proportionality constant between the critical temperature and the phase stiffness has been modified by the RG, similarly to what happens to critical exponents that dictate the scaling behavior of correlations and thermodynamic quantities near the critical point under RG rescaling. \\
Due to Eq.(\ref{bb}), the upper bound for the BKT critical temperature ($T_F/8$) that has been predicted theoretically in the context of 2D Fermi superfluids and has been used for comparison with the experiments in Refs.\cite{Nakagawa, Shi} is  different when more valleys are present in the band structure and RG effects are taken into account, that is:
\begin{equation}
    T_{BKT} \leq \frac{0.948}{4 \pi}s T_F
\end{equation}
where $T_F$ is the Fermi temperature and $s$ is the number of valleys. For Li$_x$ZrNCl is $s=2$.\\

\textit{Comparison with experiments.}---In unconventional superconductors, such as cuprates, organics, and iron-based superconductors, when carrier density is low, strong electron correlation effects and magnetic ordering can add complexity that can lead to a smearing of the BKT transition, that becomes a gradual
crossover \cite{lara}.
\begin{figure}[h!]
\centering
\includegraphics[width=.50\textwidth]{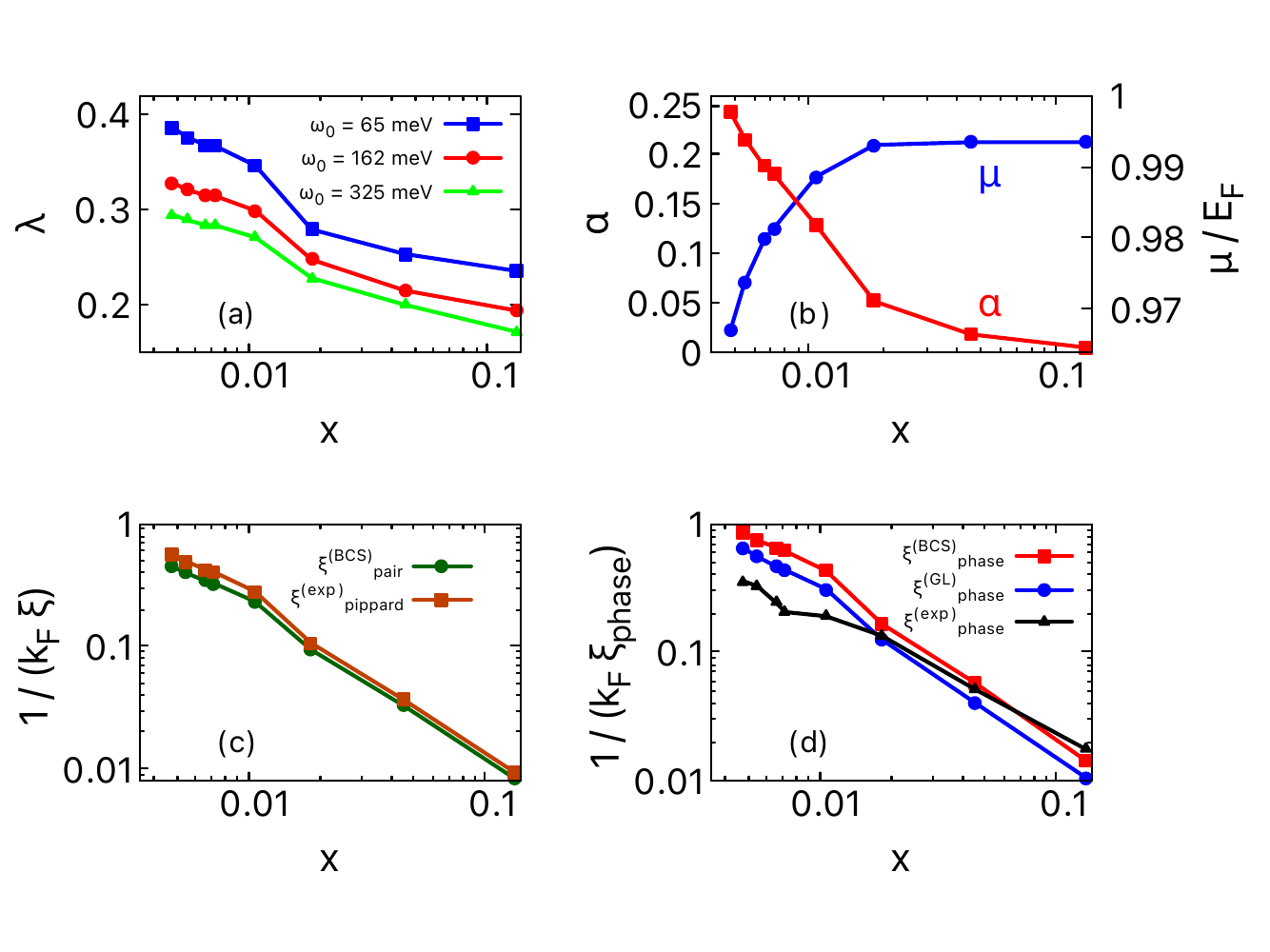}% Here is how to import EPS art
\caption{Mean-field parameters and comparison with data in Ref.\cite{Nakagawa} for the phase and the Pippard coherence lengths as functions of Lithium concentration $x$. (a) Coupling  parameter $\lambda$ for different energy cutoffs. (b) Chemical potential $\mu$ in units of the Fermi energy $E_F$ and condensate fraction. (c) Analytical intra-pair coherence length (green line) and experimental Pippard coherence length (orange line). (d) Analytical phase coherence length calculated using BCS theory (red line), GL theory (blue line) compared with data (black line). Lengths are in units of the Fermi wave-vector $k_F$. }
\label{fig4}
\end{figure}
This makes Li$_x$ZrNCl an ideal platform to study the BKT physics in the context of the BCS-BEC crossover since the undoped
state at $x = 0$ is a band insulator, free from electron correlation effects, magnetic orders, and density waves. Moreover, since the mean free path of the electrons is comparable to the Pippard coherence length $\xi_{Pippard}$, the system is far from the dirty limit.
In this material, a dimensional crossover from anisotropic 3D to 2D superconductivity occurs because of the reduced interlayer hopping in the lightly doped regime. In this regime, the superconducting BKT transition has been probed by resistivity and in-plane upper critical field measurements \cite{Nakagawa}.
\begin{figure}[h!]
\centering
\includegraphics[width=.46\textwidth]{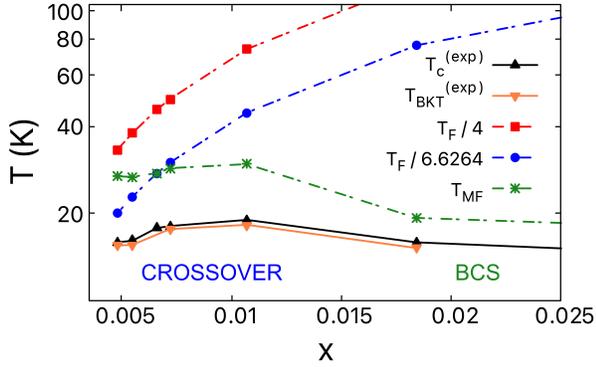}% Here is how to import EPS art
\caption{Experimental data of the critical temperature (black and orange lines) from Ref.\cite{Nakagawa} obtained with different techniques as functions of $x$. We report two upper bounds for a band with two valleys, based on the NK criterion with (blue line) and without (red line) the inclusion of RG flow equations, that are respectively $T_F / 6.6264$ amd $T_F / 4$. The mean-field critical temperature $T_{MF}$ is also shown (green line).}
\label{fig2}
\end{figure}
Fig.\ref{fig4} (a) shows $\lambda$ for different energy cutoffs. The calculated $T_{BKT}$ is not strongly affected by the choice of $\omega_0$. We take $\omega_0=65$ meV hereafter. Fig.\ref{fig4} (b) shows the condensate fraction $\alpha$, which is defined as the ratio between the number of Cooper pairs and the total number of carriers in the system, and $\mu$ in units of $E_F$ as functions of $x$. For small $x$ the system leaves the BCS regime and slightly enters the BCS-BEC crossover regime, as confirmed by $\alpha$ which becomes larger than the generally accepted value of $0.2$ as the borderline between the BCS and BCS-BEC crossover regimes \cite{Neilson2014,Rios2018}. $\mu$ is only slightly renormalized when the system moves toward the BCS-BEC crossover regime, following the BCS behavior. Fig.\ref{fig4} (c) shows the theoretical intra-pair coherence length $\xi_{Pair}$ \cite {Pisto} and the experimental Pippard coherence length $\xi^{(exp)}_{Pippard}=\hbar v_F/ \pi \Delta$, where $v_F$ is the Fermi velocity.
\begin{figure}[h!]
\centering
\includegraphics[width=.42\textwidth]{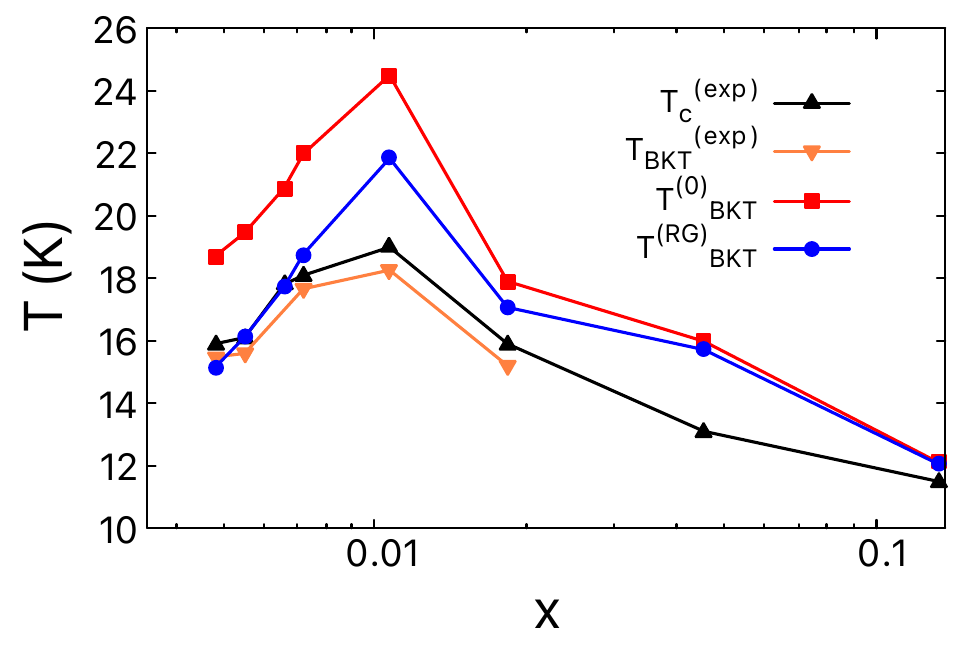}% Here is how to import EPS art
\caption{Comparison between experimental data (black and orange lines) and BKT theory for the critical temperature of Li$_x$ZrNCl as function of $x$. The red line is the BKT result without RG, while the blue line is the result including RG.}
\label{fig3}
\end{figure}
We obtain a very good agreement between theory and experiment. It is well known in fact that $\xi_{Pair}$ reduces to $\xi_{Pippard}$ in the BCS limit \cite{Pisto, Pisto1}. This is another confirmation that the system is only entering the BCS-BEC crossover regime, withouth reaching the BEC regime. Fig.\ref{fig4} (d) shows the theoretical phase coherence length $k_F \xi_{phase}$ and the experimental one as functions of $x$. We evaluate the theoretical $\xi_{phase}$  both using a Ginzburg-Landau approach \cite{Ord} and a BCS approach \cite{Pisto}. $\xi^{(exp)}_{phase}$ is extracted from the measurement of the 
\begin{figure}[h!]
\centering
\includegraphics[width=.47\textwidth]{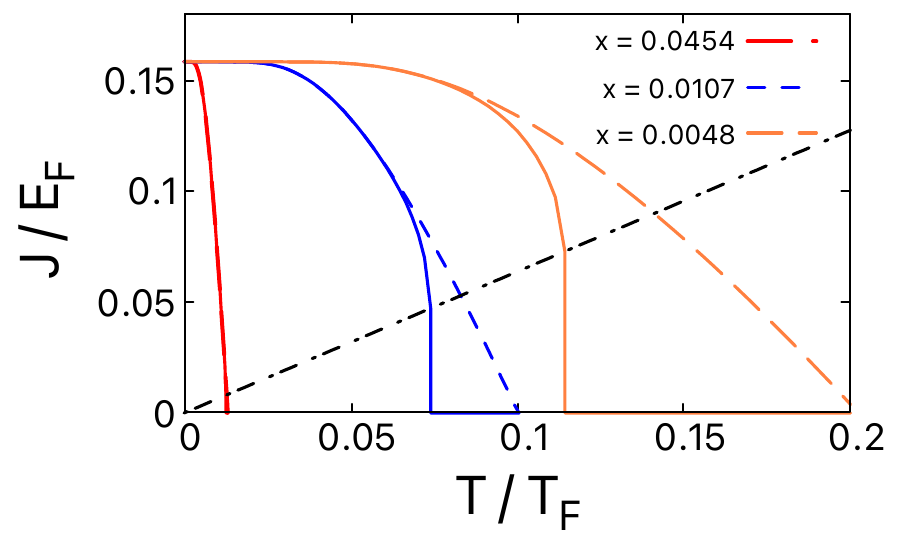}% Here is how to import EPS art
\caption{Phase stiffness $J$ in units of the Fermi energy $E_F$ as a function of temperature $T$ in units of the Fermi temperature $T_F$, for different values of  $x$. The dashed lines correspond to the bare stiffness while the solid lines are the renormalized one. The black dot dashed line corresponds to the Nelson-Kosterlitz criterion in Eq.(\ref{1t}).}
\label{fig1}
\end{figure}
temperature dependence of the upper critical field $H_{c_2}(T)$. There is good agreement between $\xi^{(exp)}_{phase}$ and the theoretical $\xi_{phase}$ in all the experimental range used for $x$, both according to BCS and Ginzburg-Landau theory. Moreover, the BCS limit $\xi^{BCS}_{phase}/\xi^{BCS}_{pair}=1/\sqrt{3}$ \cite{Pisto} is recovered theoretically and experimentally by replacing the $\xi_{Pair}$ with $\xi_{Pippard}$. In Fig. \ref{fig2} we compare the experimental data with the mean-field critical temperature and the upper bound for $T_{BKT}$ of 2D Fermi superfluids, that in the case of two valleys is $T_F/4$. The data are closer to the mean-field critical temperature $T_{MF}$ when the system is in the BCS regime for higher doping, while start to deviate from the mean-field behavior when approaching the BCS-BEC crossover regime. However, since the system is only entering the crossover regime, the upper bound to the critical temperature is far from the experimental results, since this limit is reached only in the BEC regime. However, this estimation is not sufficient to reproduce the experimental result, but gives a good qualitative description of the non-monotonic behavior of the critical temperature as a function of doping, which follows the mean-field behavior in the BCS regime, while in the crossover regime the doping dependence is described by the BKT physics.

In Fig. \ref{fig3}, we compare our theoretical results for $T_{BKT}$, to experimental outcomes for Li$_x$ZrNCl \cite{Nakagawa}, both with and without the inclusion of the RG. We note that there is general good agreement between theory and experiments, with our theoretical model that succeed to capture the behavior of the experimental transition temperatures measured with different techniques as a function of $x$, in particular the position of the maximum of $T_{BKT}$ and the inflection point, a feature that is not correctly described in Ref. \cite{Shi}. Furthermore, it is clear that the RG is very important to obtain a correct quantitative estimation, especially when the system is in the BCS-BEC crossover regime. In Fig.\ref{fig1} we report the evolution of the renormalized and bare phase stiffness as functions of temperature, for three values of $x$, that we use as a prediction for Li$_x$ZrNCl, given that in Ref.\cite{Nakagawa} this quantity has not been measured. Since this material is in the clean limit for electronic transport, we suggest that the sharp drop of the phase stiffness predicted by the BKT theory can be observed in this material, not only in the BCS regime as observed in NbN \cite{Weitzel}, but across the BCS-BEC crossover where the deviation from mean field occurs in a broader range of temperatures measured from $T_{BKT}$. For the considered values of $x$, we predict a deviation from the mean-field behavior of the renormalized stiffness occuring in a temperature range of $0.4$ K, $3.3$ K and $4.2$ K for $x=0.0454$, $x=0.0107$, $x=0.0048$ respectively, where the renormalized stiffness scales as a square root close to the jump at $T_{BKT}$, in contrast with the BCS linear behavior approaching $T_{MF}$. The renormalization of the phase stiffness, larger for lower levels of doping, further suggests that the system is entering the BCS-BEC crossover regime without reaching BEC. In fact, in Ref.\cite{Bighin} is shown how the renormalization in the BEC regime becomes again smaller than in the BCS-BEC crossover. This behavior could be observed in Li$_x$ZrNCl when lower dopings will be experimentally accessible, making possible for the system to reach the BEC regime.\\

\textit{Conclusions.}---In this work, we demonstrate the predictive power of the BKT theory for the evaluation of the critical temperature of the superconducting transition in 2D systems. The inclusion of the RG flow equations turns out to be crucial to obtain a satisfactory quantitative agreement with experiments, especially when the system is driven toward the crossover regime. Therefore, as opposite to other microscopic theory of superconductivity, as the BCS or the Eliashberg theory, for which the prediction of the critical temperature is not practically possible, in particular in the crossover regime of the BCS-BEC crossover, where pair fluctuations suppress the critical temperature in a complicated way, the BKT theory has predictive power for the evaluation of the critical temperature throughout the BCS-BEC crossover. Moreover, since Li$_x$ZrNCl is a clean material, a sharp drop in the phase stiffness at the BKT transition as the one observed in recent experiments on NbN can be observed not only in the BCS regime, corresponding to NbN, but also across the BCS-BEC crossover for the first time in a solid state system by tuning the Lithium concentration. In this way the deviation from the mean-field expectation for the phase stiffness caused by RG can be observed in a broader range of temperature, with the specific behavior predicted in this work. Our approach can represent a powerful tool for the prediction of the phase stiffness and $T_{BKT}$ of novel 2D superconductors and superfluids \cite{Midei1,Midei2,Paramasivam}, once the microscopic parameters of their ground state are known.
\\
\\
We acknowledge Roberta Citro, Sebastiano Pilati, and Giulia Venditti for useful discussions. This work has been supported by PNRR MUR project PE0000023-NQSTI. KF and LS are partially supported by “Iniziativa Specifica Quantum” of INFN,
by PRIN Project “Quantum Atomic Mixtures: Droplets, Topological Structures, and
Vortices” of the Italian Ministry for Universities and Research. KF and LS
acknowledge the Project ”Frontiere Quantistiche” (Dipartimenti di Eccellenza)
of the Italian Ministry for Universities and Research. LS is partially supported by the
European Quantum Flagship Project “PASQuanS 2”, by the NextGeneration-EU within
the National Center for HPC, Big Data and Quantum Computing
[Project No.5CN00000013, CN1 Spoke 10: “Quantum Computing”], and
by the BIRD Project “Ultracold atoms in curved geometries” of the University of Padova.

\end{document}